# Photometric and Spectroscopic Observations of Comet 2P/Encke's Nucleus in the Mid-Infrared Wavelength Region

Takafumi Ootsubo[1], Hideyo Kawakita[2], Jun-ichi Watanabe[2,3], Hitomi Kobayashi[2,4], Yoshiharu Shinnaka[2]

[1] Department of Basic Sciences, University of Occupational and Environmental Health, Japan, 1-1, Iseigaoka, Yahatanishi-ku Kitakyushu, Fukuoka, 807-8555, Japan

[2] Koyama Space Science Institute, Kyoto Sangyo University, Kamigamo Motoyama, Kita-ku, Kyoto, Kyoto 603-8555, Japan

[3] National Astronomical Observatory of Japan, 2-21-1 Osawa, Mitaka, Tokyo 181-8588, Japan

[4] Photocross Co. Ltd., 17-203 Iwakura-Minami-Osagicho, Sakyo-ku, Kyoto, Kyoto 606-0003, Japan

ABSTRACT

We observed comet 2P/Encke with the Cooled Mid-Infrared Camera and Spectrometer (COMICS) on the Subaru telescope on UT 2003 November 12.3, when the comet was at heliocentric and geocentric distances of 1.09 and 0.27 au, respectively. Imaging observations of the comet with narrow-band filters centered at 8.8, 10.5, 11.7, 12.4, 18.8, and 24.5 $\mu$m and low-resolution spectroscopic observations in the *N*-band (8–13 $\mu$m, R~250) of the comet were performed.

The observed spatial profiles of the comet were consistent with those of a standard star (FWHM ~ $0''.3$ at 10.5 $\mu$m, close to the diffraction-limited image size), indicating that the observed spectrum was dominated by thermal emission from the cometary nucleus. The semi-axial ratio ($a/b$) of the nucleus of 2P/Encke is estimated as $2.26^{+0.33}_{-0.21}$ based on the isothermal, prolate ellipsoid model. The photometric data with the correction for rotational phases, can be explained by a blackbody temperature of 328 ± 7 K, consistent with the nucleus surface temperature empirically estimated at 1.09 au from the Sun (327 K). This blackbody approximation for the nucleus leads to $a = 3.6^{+0.5}_{-0.3}$ km and $b = 1.6^{+0.2}_{-0.2}$ km by assuming an emissivity of unity.

The observed low-resolution *N*-band spectrum showed an overall smooth continuum with no prominent excess emission feature at ~10 $\mu$m due to fine silicate grains in the cometary coma. The relative emissivity spectrum of the nucleus showed a negative feature at ~9.5 $\mu$m, as seen in the mid-infrared nucleus spectrum of 460P/PanSTARRS [T. Ootsubo, et al. (2021), Icarus, 363, 114425].

## 1. Introduction

Cometary nuclei are considered remnants of planetesimals from the early Solar System. The shape, size, and interior structure (e.g., density distribution and chemical composition of ices) of each nucleus likely reflect not only the initial formation processes of the planetesimals 4.6 Gyrs ago, but also the subsequent erosion and fragmentation of the cometary nuclei during orbits with perihelion distances of a few au. Therefore, understanding these processes through observations of cometary nuclei will provide a clearer view of the initial conditions of planetesimal formation in the Solar System. Generally, however, their small size makes it difficult to resolve them using ground-based facilities or space telescopes. Direct imaging (and probing the interior) of a cometary nucleus is typically only possible from a spacecraft operating in close proximity (e.g., the Giotto flyby of 1P/Halley and the Rosetta rendezvous with 67P/Churyumov-Gerasimenko). Alternatively, ground-based radar observations can reveal the shape and size of the nucleus, but only for comets making very close approaches to Earth.

Although it is difficult to directly observe the shape of a cometary nucleus, its physical properties (such as effective radius, geometric albedo, spin rate, and semi-axial ratios assuming an ellipsoidal shape) can be derived from remote observations (M. M. Knight et al. 2024). A cometary nucleus is typically difficult to resolve from ground-based observatories, as it is obscured by an optically thick dust coma when the comet is within ~3 au of the Sun. Conversely, for comets at larger heliocentric distances or those that are intrinsically inactive, the coma contribution is either negligible or can be easily subtracted, making it much easier to extract the nuclear signal even using ground-based facilities (T. Ootsubo et al. 2026). Even in mid-infrared observations, the thermal emission from cometary dust grains in the coma usually dominates, making it nearly impossible to distinguish the nuclear emission from that of the dust (C. G. Mason et al. 2001). However, for a comet exhibiting very low activity even at ~1 au from the Sun (e.g., 460P/PanSTARRS, also known by its provisional designation P/2016 $BA_{14}$), thermal emission directly from the nucleus can be successfully detected (T. Ootsubo et al. 2021).

Here we present measurements of the thermal emission from the nucleus of comet 2P/Encke at 1.09 au from the Sun (on UT 2003 November 12). Although 2P/Encke was active during our observations (~$4 \times 10^{27}$ molecules $s^{-1}$ on UT 2003 November 4–6; Y. L. Radeva et al. 2013; M. Lippi et al. 2021), we detected the nucleus of this comet in the mid-infrared wavelength region, likely due to its low dust-to-gas ratio and the high spatial resolution of our observations. 2P/Encke is a short-period comet with an orbital period of 3.3 years. Its Tisserand invariant with respect to Jupiter is $T_J = 3.025$, indicating that its orbit lies entirely inside that of Jupiter (with an aphelion distance of 4.09 au). Studying the nuclear evolution of short-period comets, such as Jupiter-family comets (JFCs), is of fundamental importance because their frequent perihelion passages accelerate their physico-chemical evolution. Although 2P/Encke is dynamically classified as an Encke-type comet rather than a JFC (H. F. Levison & L. Dones 2014), it serves as a critical evolutionary analog. Ultimately, to constrain the initial conditions of planetesimals in the protoplanetary disk, we must first understand how these cometary nuclei evolve over time. The physical and surface properties of cometary nuclei derived from remote observations, including those of JFCs and other dynamical classes, are comprehensively reviewed by M. M. Knight et al. (2024).

The physical properties of 2P/Encke's nucleus have been extensively studied across a wide range of wavelengths, from the optical to the radio (Z. Sekanina 1988a, 1988b, 1991; M. C. Festou & O. Barale 2000; Y. R. Fernández et al. 2000, 2005; M. J. S. Belton et al. 2005; J. K. Harmon & M. C. Nolan 2005; M. S. Kelley et al. 2006; S. C. Lowry & P. R. Weissman 2007; K. Jockers et al. 2011; and references therein). Z. Sekanina (1988a, 1988b, 1991) analyzed the rotation pole of 2P/Encke's nucleus based on observations of its fan-shaped coma during the apparitions from 1868 to 1984. M. C. Festou & O. Barale (2000) reanalyzed the data of 2P/Encke in 1980 apparition and found similar solution of the rotation axis with an accuracy of a few degrees. K. Jockers et al. (2011) also investigated the asymmetric coma of CN in 2P/Encke and derived the direction of its spin axis. All of these studies yielded consistent results, placing the pole at approximately $(\mathrm{RA2000}, \mathrm{Dec2000}) = (200°, 0°)$. Furthermore, M. J. S. Belton et al. (2005) suggested that the nucleus is likely in either a relatively low-excitation short-axis mode (SAM) or a state of principal-axis rotation, based on their analysis of previous photometric observations.

J. K. Harmon & M. C. Nolan (2005) reported their radar observations of 2P/Encke during the close approach to Earth in November 2003. They argued that the nucleus of 2P/Encke is an oblong object with a long-axis dimension of ~9 km. If the nucleus is assumed to be a prolate spheroid with semi-axial ratio $a/b$ (where $a$ and $b$ are the semi-major and semi-minor axes, respectively), the semi-axial ratio is estimated as $a/b = 2.04$ by M. J. S. Belton et al. (2005) based on their reinterpretation of the infrared data obtained by Y. R. Fernández et al. (2000). The effective nucleus radius and the visual geometric albedo of 2P/Encke are estimated as 2.4 ± 0.3 km and 0.05 ± 0.02, respectively (Y. R. Fernández et al. 2000). S. C. Lowry & P. R. Weissman (2007) reported the rotational period of Encke's nucleus to be 11.083 ± 0.003 h and derived $a/b \geq 1.44 \pm 0.06$ for the axial ratio of Encke's nucleus based on the light curves of 2P/Encke in the optical region. K. Jockers et al. (2011) also discussed the rotational period of Encke's nucleus based on the morphology of CN coma and concluded that the rotational period was 11.056 ± 0.024 h in November 2003.

Photometric and spectroscopic observations of 2P/Encke in mid-infrared wavelength region were performed using the Spitzer Space Telescope in November 2003 and in June 2004 (M. S. Kelley et al. 2006). The effective radius of the nucleus was derived as 1.72 ± 0.10 km from the 2003 November observations, and 2.34 ± 0.14 km from the 2004 June observations. These results may indicate $a/b \geq 1.4$ for the semi-axial ratio of Encke's nucleus. Their spectra, observed at 2.6 au from the Sun, did not exhibit significant silicate features at ~10 $\mu$m and ~20 $\mu$m.

2. Observations

Comet 2P/Encke was observed on UT 2003 November 12.3 using the Cooled Mid-Infrared Camera and Spectrometer (COMICS) (H. Kataza et al. 2000; Y. Okamoto et al. 2003) mounted on the 8.2-m Subaru telescope atop Maunakea, Hawaii. Diffraction-limited spatial resolution was achieved for all observing modes of the COMICS. Our imaging observations in N- and Q-bands together with low-resolution spectroscopic observations in N-band were carried out on UT 2003 Nov 12.3, when the comet was at heliocentric and geocentric distances of 1.09 and 0.27 au, respectively. The phase angle of 2P/Encke was 62° on that night.

Imaging observations of the comet were conducted using narrow-band filters at 8.8 $\mu$m ($\Delta\lambda$ = 0.8 $\mu$m), 10.5 $\mu$m ($\Delta\lambda$ = 1.0 $\mu$m), 11.7 $\mu$m ($\Delta\lambda$ = 1.0 $\mu$m), 12.4 $\mu$m ($\Delta\lambda$ = 1.2 $\mu$m), 18.8 $\mu$m ($\Delta\lambda$ = 0.9 $\mu$m), and 24.5 $\mu$m ($\Delta\lambda$ = 1.9 $\mu$m). The field of view for imaging observations is $42'' \times 32''$ with a plate scale of $0''.13$ pixel$^{-1}$. Additionally, low-resolution spectroscopic observations of the comet were performed in the *N*-band (8–13 $\mu$m, $R = \lambda/\Delta\lambda \sim 250$) with a slit ($0.''33 \times 40''$ with a plate scale of $0''.165$ pixel$^{-1}$). To cancel high thermal background radiation from the telluric atmosphere, secondary mirror chopping was used at a frequency of 0.45 Hz with an amplitude of 20". We selected HD 197989 and HD 203504 as flux standard stars for the imaging observations (M. Cohen et al. 1999). The absolute flux values of these standard stars for the filters were obtained by integrating the template spectra provided by M. Cohen et al. (1999). We also observed HD 203504 as a spectrophotometric standard star for the low-resolution spectroscopy. The wavelength calibration was performed using the sky emission lines. The observational conditions of the comet and the standard stars are listed in Table 1.

The data reduction was performed by using the IRAF software (Tody 1986, 1993) and the specialized tools provided by the COMICS instrument team (https://subarutelescope.org/Observing/DataReduction/index.html). The aperture size for the photometry was set to $1''.3$ in radius. The spectra were extracted with a slit region of $0.''33$ (slit width) by $0.''83$ (along the slit direction) to achieve an adequate signal-to-noise ratio of the extracted spectrum. Slit-loss corrections for the spectra were adopted based on the photometric data (Y. Shinnaka et al. 2018 for more detail).

The sky conditions were basically clear throughout our observations. We confirmed that the observed fluxes of the standard stars did not show significant short-term variations, indicating that the photometric calibrations were not severely affected by thin clouds. Although the variable wind conditions might have affected the telescope pointing stability and tracking, potentially causing some slit losses during the low-resolution spectroscopic observations, the absolute flux level of the extracted spectrum was accurately calibrated by scaling it to match the photometric data. Therefore, the sky conditions did not significantly impact our final results.

Table 1: Observational circumstances.

| UT Date | UT Time (at start) | Object | Integration time [s] | Observing mode (*) | Airmass |
|---|---|---|---|---|---|
| 2003 Nov 12 | 04:31 | HD 197989 | 10.6 | N12.4 | 1.05 |
| | 04:34 | HD 197989 | 10.8 | N8.8 | 1.05 |
| | 04:38 | HD 197989 | 21.6 | Q18.8 | 1.05 |
| | 04:43 | HD 197989 | 182.8 | Q24.5 | 1.06 |
| | 05:05 | 2P/Encke | 21.7 | N8.8 | 1.06 |
| | 05:08 | 2P/Encke | 21.2 | N12.4 | 1.07 |
| | 05:12 | 2P/Encke | 41.4 | Q18.8 | 1.07 |
| | 05:16 | 2P/Encke | 300.7 | Q24.5 | 1.08 |
| | 05:32 | 2P/Encke | 21.2 | N11.7 | 1.09 |
| | 05:37 | 2P/Encke | 21.2 | N10.5 | 1.09 |
| | 05:59 | HD 203504 | 21.2 | N10.5 | 1.09 |
| | 06:01 | HD 203504 | 21.2 | N11.7 | 1.09 |
| | 06:04 | HD 203504 | 21.2 | N12.4 | 1.10 |
| | 06:06 | HD 203504 | 21.7 | N8.8 | 1.10 |
| | 06:10 | HD 203504 | 122.1 | Q18.8 | 1.11 |
| | 06:30 | HD 203504 | 43.4 | NL Spec | 1.17 |
| | 06:53 | 2P/Encke | 180.8 | NL Spec | 1.26 |
| | 07:09 | HD 203504 | 43.4 | NL Spec | 1.30 |
| | 07:13 | HD 203504 | 21.7 | N8.8 | 1.32 |
| | 07:15 | HD 203504 | 21.2 | N10.5 | 1.33 |
| | 07:18 | HD 203504 | 21.2 | N11.7 | 1.34 |
| | 07:20 | HD 203504 | 21.2 | N12.4 | 1.35 |
| | 07:24 | HD 203504 | 43.3 | Q18.8 | 1.37 |
| | 07:42 | 2P/Encke | 82.7 | Q18.8 | 1.46 |
| | 07:49 | 2P/Encke | 21.2 | N12.4 | 1.49 |
| | 07:51 | 2P/Encke | 21.2 | N11.7 | 1.51 |
| | 07:53 | 2P/Encke | 21.2 | N10.5 | 1.52 |
| | 07:55 | 2P/Encke | 21.7 | N8.8 | 1.53 |

(*) "N8.8" means the imaging mode with the filter at 8.8 $\mu$m in *N*-band (others are expressed in similar manner) while "NL Spec" means the low-resolution spectroscopic mode in N-band.

## 3. Results and Analysis

### 3.1 Photometric data

Figure 1 shows an example image of 2P/Encke taken by the COMICS. By subtracting paired images from each other (for chopping beams), the thermal background radiation was canceled, and two signals (positive and negative) were obtained. Comet 2P/Encke showed no clear coma in any of the images taken, regardless of the filter used, and always appeared as a point source. Figure 2 shows the radial intensity profiles of 2P/Encke compared with those of a standard star (as a point source) for the same filter. These profiles indicate that the observed fluxes of 2P/Encke were dominated by the thermal radiation from a very compact source (at least smaller than FWHM ~ 0".3 of the point spread function (PSF) at 10.5 $\mu$m, corresponding to ~60 km in diameter at the comet). It is likely that these signals originate from the thermal radiation of the cometary nucleus.

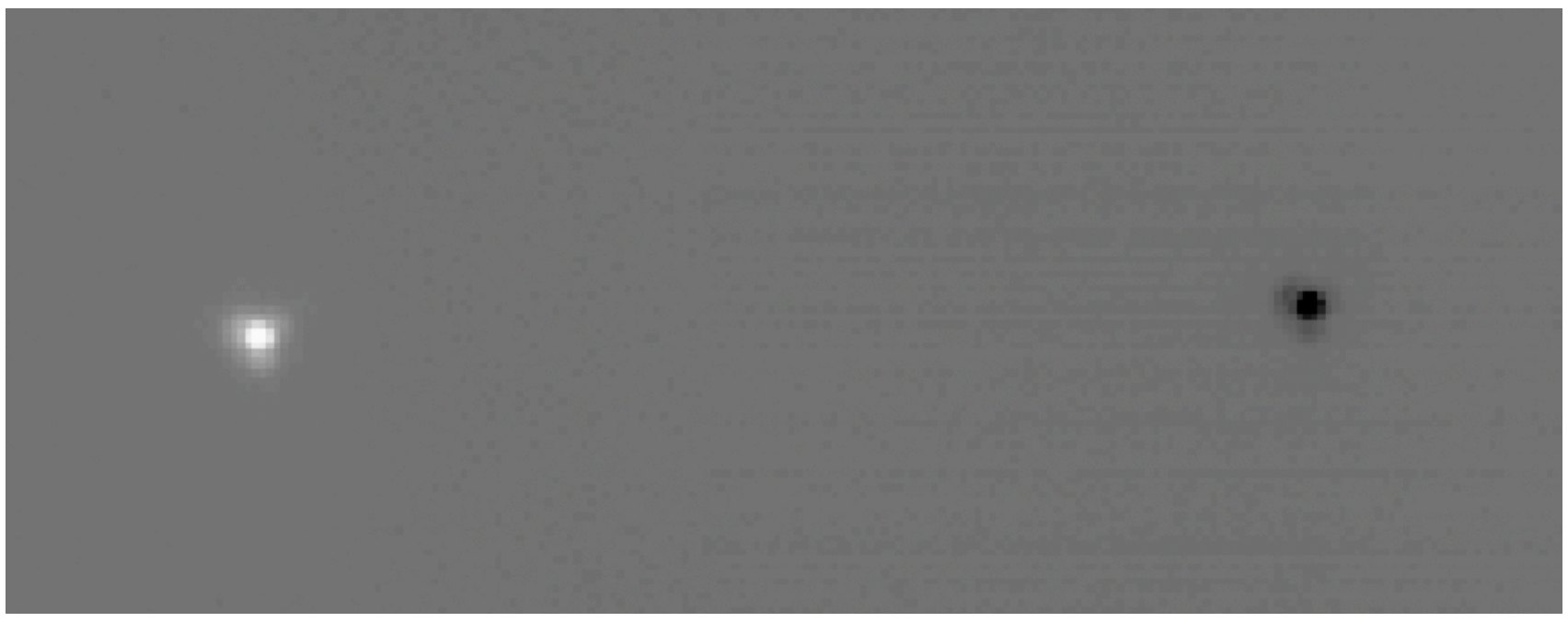

Figure 1: An example of imaging observation of 2P/Encke with COMICS at 10.5 $\mu$m. Due to the 20" chopping amplitude, both positive and negative images are recognized after the thermal background subtraction. The diffuse extensions around the bright spots correspond to the diffraction pattern of the Subaru Telescope.

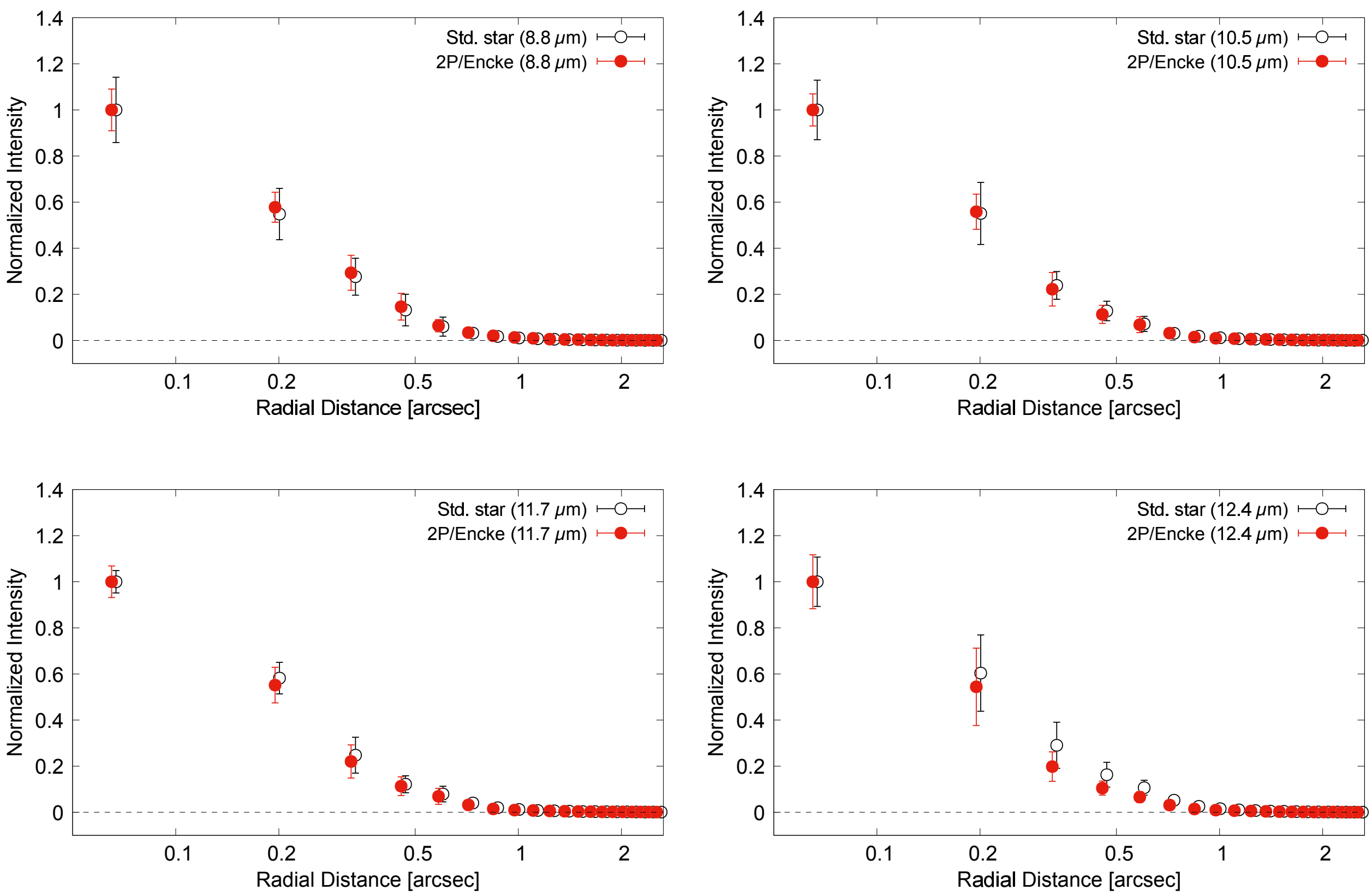

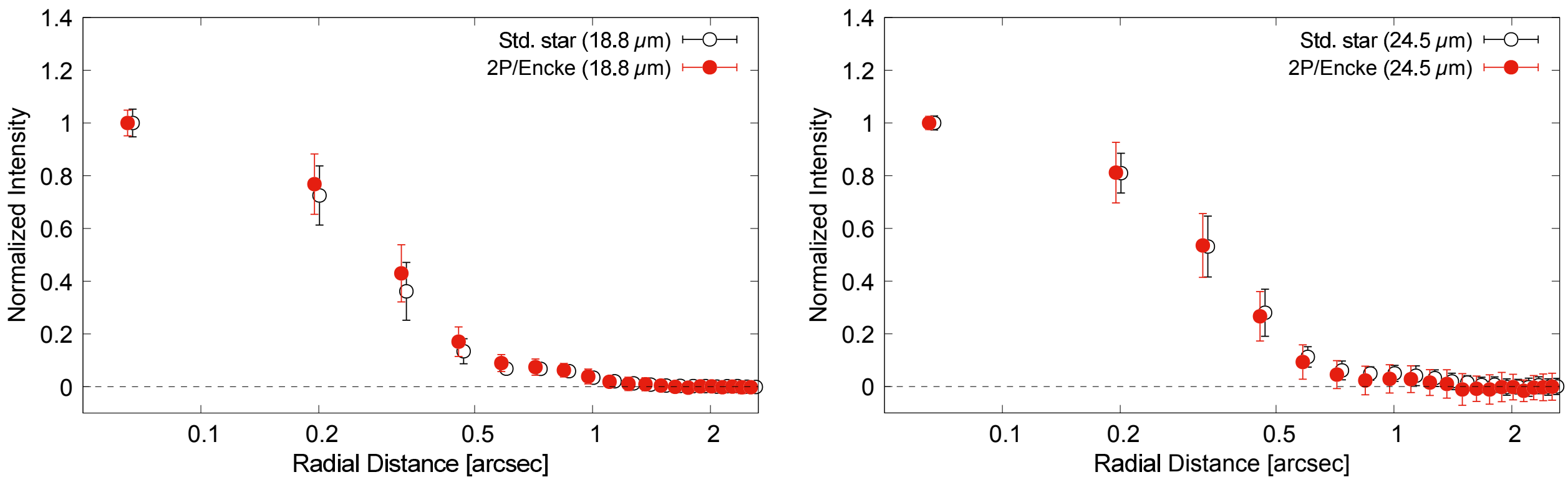

Figure 2: Radial intensity profiles of 2P/Encke (red filled circles) compared with those of a standard star (black open circles), showing consistency with a point source in all filters. Note that the spatial profiles are consistent between the comet and the standard star across all filters, indicating that the observations were nearly diffraction-limited and any issues of varying seeing are negligible.

Table 2 lists the results of our photometric observations of 2P/Encke. Note that the flux density of 2P/Encke at 8.8 $\mu$m at the end of our observations (at 07:56 UT) was larger by a factor of 1.8 than the flux density observed at the beginning of the observations (at 05:05 UT). Figure 3 plots the observed flux densities in the N-band. The "early" data set (05:05–05:38 UT on 2003 November 12) is systematically fainter than the "late" data set (07:42–07:56 UT on 2003 November 12) by a factor of 1.5–1.8 for each narrow-band filter. Because the observed fluxes were dominated by the thermal radiation of the cometary nucleus as argued above, this change in the observed flux may be explained by the rotation of the nucleus.

Table 2: Photometry of 2P/Encke

| UT Time (+) on 2003 Nov 12 | Observing mode | Band flux density [W/m$^2$/$\mu$m] |
|---|---|---|
| 05:05 | N8.8 | $(9.17 \pm 0.25) \times 10^{-14}$ |
| 05:09 | N12.4 | $(7.60 \pm 0.21) \times 10^{-14}$ |
| 05:13 | Q18.8 | $(3.60 \pm 0.14) \times 10^{-14}$ |
| 05:23 | Q24.5 | $(1.51 \pm 0.08) \times 10^{-14}$ |
| 05:32 | N11.7 | $(9.20 \pm 0.25) \times 10^{-14}$ |
| 05:38 | N10.5 | $(11.1 \pm 0.30) \times 10^{-14}$ |
| 07:44 | Q18.8 | $(6.52 \pm 0.29) \times 10^{-14}$ |
| 07:49 | N12.4 | $(13.1 \pm 0.36) \times 10^{-14}$ |
| 07:51 | N11.7 | $(13.8 \pm 0.37) \times 10^{-14}$ |
| 07:53 | N10.5 | $(16.5 \pm 0.45) \times 10^{-14}$ |
| 07:56 | N8.8 | $(16.3 \pm 0.44) \times 10^{-14}$ |

(+) The midpoint of the total exposure duration.

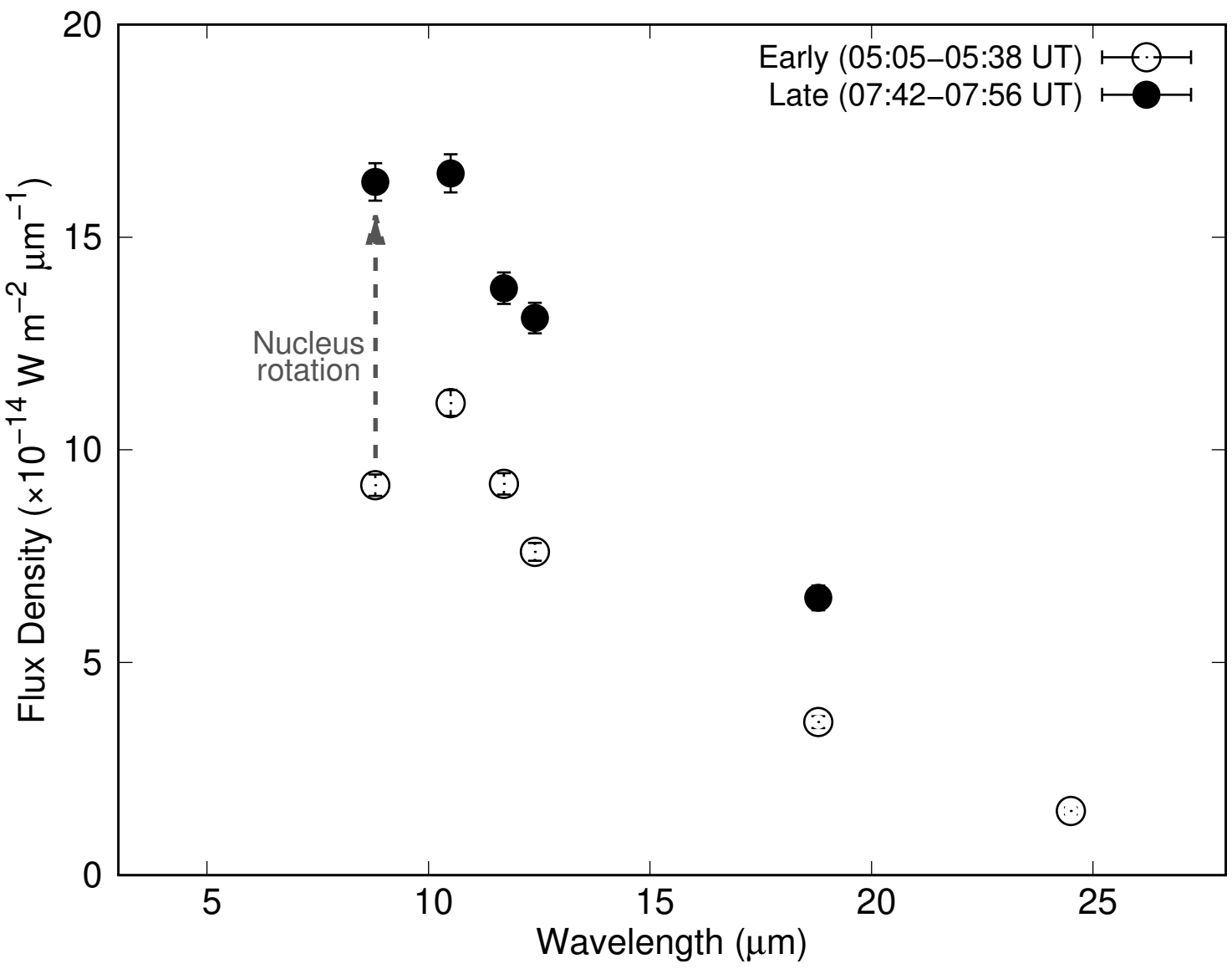

Figure 3: Photometric results of 2P/Encke. Open circles represent the "early" observations from 05:05 to 05:38 (UT) on 2003 Nov 12 while filled circles represent the "late" observations from 07:42 to 07:56 (UT) on the same date. The difference between these two sets is considered due to nucleus rotation of 2P/Encke.

The rotational period of the nucleus of 2P/Encke was estimated as 11.083±0.003 h based on photometric results of 2P/Encke in the optical wavelength range (S. C. Lowry & P. R. Weissman 2007). The time difference between the first and the last observations of N8.8 is 2.85 h, close to a quarter of the rotational period of the nucleus. If the nucleus is elongated and rotating around an axis perpendicular to the elongated axis of the nucleus, the cross-section of the nucleus viewed from the Earth may change significantly, especially in the case that the spin axis is perpendicular to the line-of-sight. Thus, if we assume (i) a single-axis rotation of the nucleus, (ii) a uniform temperature and emissivity on the nucleus surface, and (iii) a prolate nucleus shape, the observed change in flux density simply depends on the cross-section of the nucleus. The semi-axial ratio ($a/b$) of the nucleus of 2P/Encke is expected to be 1.7–1.8, or larger, based on the ratios of flux densities at the same wavelength (at 8.8, 12.4 and 18.8 $\mu$m). This is consistent with the previous estimate, $a/b \geq 1.44 \pm 0.06$ (S. C. Lowry & P. R. Weissman 2007).

To place a more quantitative constraint on the semi-axial ratio, we performed an analysis based on the projected cross-sectional area of a rotating prolate ellipsoid, following the formalism of P. L. Lamy et al. (2004). The projected cross-section $S$ of a prolate spheroid ($a > b = c$), observed at aspect angle $\alpha$ (the angle between the spin axis and the direction to the observer; Appendix A) and at rotational angle $\phi$, is given by

$$S(\phi, \alpha) = \pi a b^2 \sqrt{\left(\frac{\sin^2\phi}{a^2} + \frac{\cos^2\phi}{b^2}\right)\sin^2\alpha + \frac{\cos^2\alpha}{b^2}} \ .$$

For thermal emission dominated by the nucleus, the observed flux density at a given wavelength is proportional to $S$, under the assumption of uniform surface temperature and emissivity.

Because the COMICS observations at different wavelengths were obtained sequentially rather than simultaneously (Table 2), the early and late observations of the same filter sample different rotational phases. The time intervals between paired observations of the same filter range from 2.25 h (N10.5) to 2.85 h (N8.8), corresponding to $73° - 93°$ of rotational phase (using the rotation period of $11.083 \pm 0.003$ h from S. C. Lowry & P. R. Weissman 2007). This naturally produces different flux ratios for different filters, as summarized in Table 3.

Table 3: Flux ratios between "early" and "late" observations of 2P/Encke.

| Observing mode | Time difference [h] | Phase difference [degrees] | Flux ratio between "early" and "late" observations: $F_{late}/F_{early}$ |
|---|---|---|---|
| N8.8 | 2.85 | 92.6 | $1.78 \pm 0.07$ |
| N10.5 | 2.25 | 73.1 | $1.49 \pm 0.06$ |
| N11.7 | 2.32 | 75.3 | $1.50 \pm 0.06$ |
| N12.4 | 2.67 | 86.6 | $1.72 \pm 0.07$ |
| Q18.8 | 2.52 | 81.7 | $1.81 \pm 0.11$ |

To apply this formalism, we require the aspect angle $\alpha$ during the COMICS observations. The geocentric position of 2P/Encke on UT 2003 November 12.27, which corresponds to the midpoint of the COMICS imaging observations and is close in time to the spectroscopic observation, was obtained from the JPL Horizons ephemeris service: RA = 21h 24m 34.0s, Dec = $+37° \, 31' \, 28''$ (J2000). Using the spin-axis direction, $(\mathrm{RA2000}, \mathrm{Dec2000}) = (204.4°, +3.9°)$ (K. Jockers et al. 2011), we obtain $\alpha = 71.7°$. This is considerably smaller than $\alpha \sim 87° - 91°$ during the Arecibo radar observations of J. K. Harmon & M. C. Nolan (2005) on November 23–25, because the comet's geocentric position shifted by approximately 35° in right ascension and 20° in declination between November 12 and November 23 due to its rapid apparent motion near closest approach on November 16. The aspect angle computed using the pole direction of Z. Sekanina (1988), $(\mathrm{RA2000}, \mathrm{Dec2000}) = (206°, +4°)$ gives a similar value of α = 72.9°.

In addition to the aspect angle, we need to determine the rotational phase at the time of each COMICS observation. For the rotation period, we adopt the synodic period of $11.083 \pm 0.003$ h derived by S. C. Lowry & P. R. Weissman (2007), who combined their October 2002 photometry with the data of Y. R. Fernández et al. (2005). For the rotation epoch, we use the result of K. Jockers et al. (2011), who determined a well-defined epoch $t_0$ = November 19.4627 UT from CN imaging and HCN spectroscopy of 2P/Encke obtained from November 19 to 25, 2003, only 7 days after our COMICS observations. This epoch is defined as the time at which the jet position angle crosses the rotation axis position angle from south to north. Using the Lowry & Weissman period with the Jockers et al. epoch, the accumulated phase uncertainty over the approximately 174 hours separating our COMICS observations from the epoch is $\pm 0.004$ rotations ($\pm 1.4°$).

We simultaneously fitted the flux ratios of all five bands (Table 3) with the prolate ellipsoid model, with the semi-axial ratio $a/b$ and the initial rotational angle $\phi_0$ at the time of the first observation (N8.8 at 05:05 UT) as free parameters. Using the aspect angle $\alpha = 71.7°$ (K. Jockers et al. 2011 pole), the best-fit parameters are $a/b = 2.26^{+0.33}_{-0.21}$ and $\phi_0 = 102 \pm 4°$, with a reduced chi-squared of $\chi^2_\nu = 0.40$ (a degree of freedom; $\nu = 3$). All five observed flux ratios are reproduced within 1-$\sigma$ of the measurement uncertainties (Figure 4). The aspect angle derived from the pole direction of Z. Sekanina (1988) gives the semi-axial ratio of $a/b = 2.19^{+0.29}_{-0.19}$.

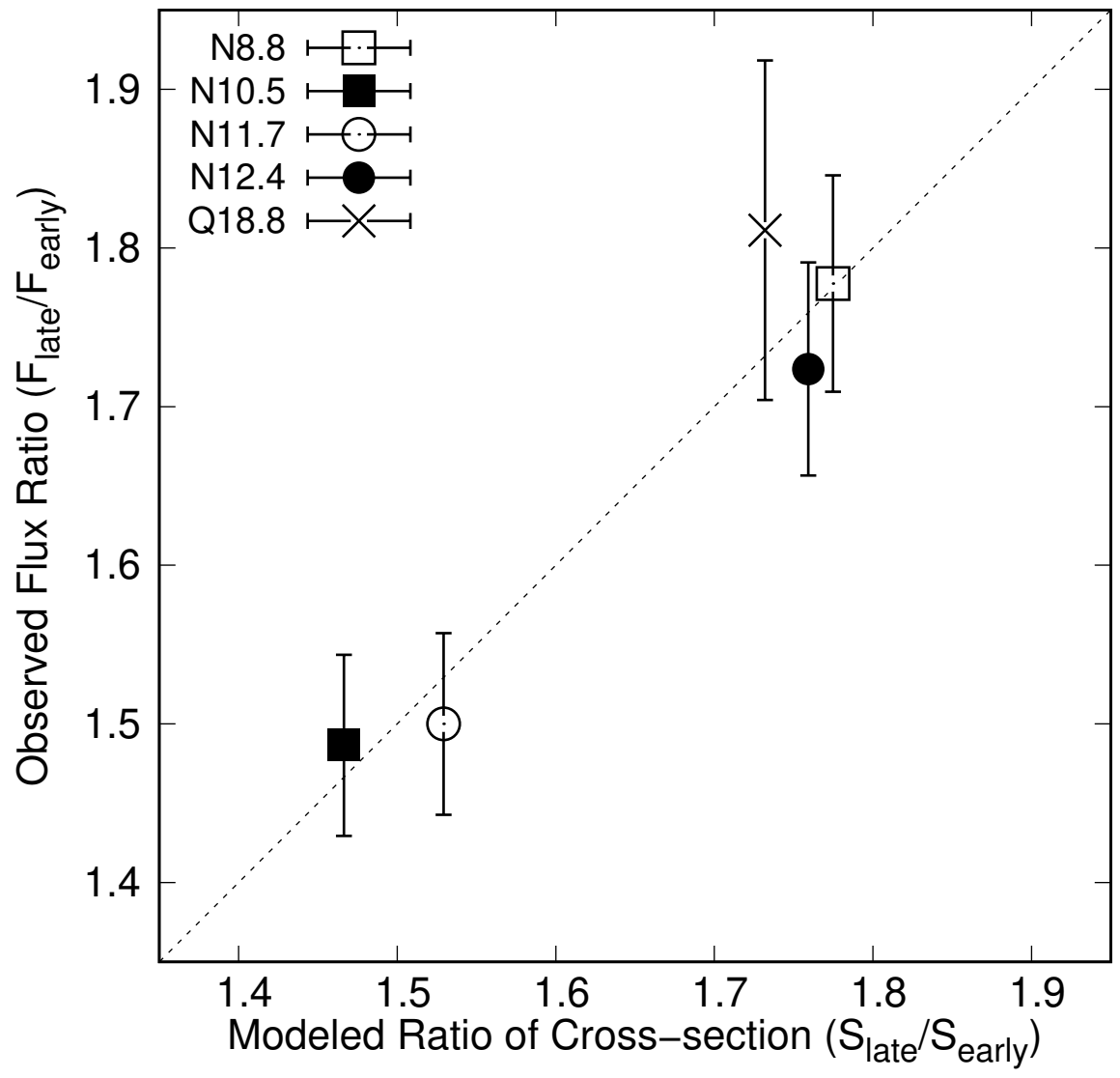

Figure 4: Observed flux ratios between "early" and "late" observations ($F_{\mathrm{late}}/F_{\mathrm{early}}$) compared with the modeled ratios of cross-section ($S_{\mathrm{late}}/S_{\mathrm{early}}$) based on the prolate ellipsoid model using the aspect angle $\alpha = 71.7°$ (K. Jockers et al. 2011 pole). The best-fit parameters are $a/b = 2.26$ and $\phi_0 = 102°$ (see text).

Using the best-fit ellipsoid model ($a/b = 2.26$, $\alpha = 71.7°$), we applied cross-section correction factors to normalize all flux measurements to the rotation phase at the time of the *N*-band spectroscopic observation (November 12.29 UT). The corrected spectral energy distribution (in Figure 5) shows that the flux densities from the early and late imaging observations are brought into mutual agreement, confirming that the observed flux variations are primarily due to the changing projected cross-section of the nucleus as it rotates. The best-fit blackbody temperature is 328±7 K. Based on this blackbody approximation, we finally estimated $a = 3.6^{+0.5}_{-0.3}$ km and $b = 1.6^{+0.2}_{-0.2}$ km by assuming an emissivity of unity.

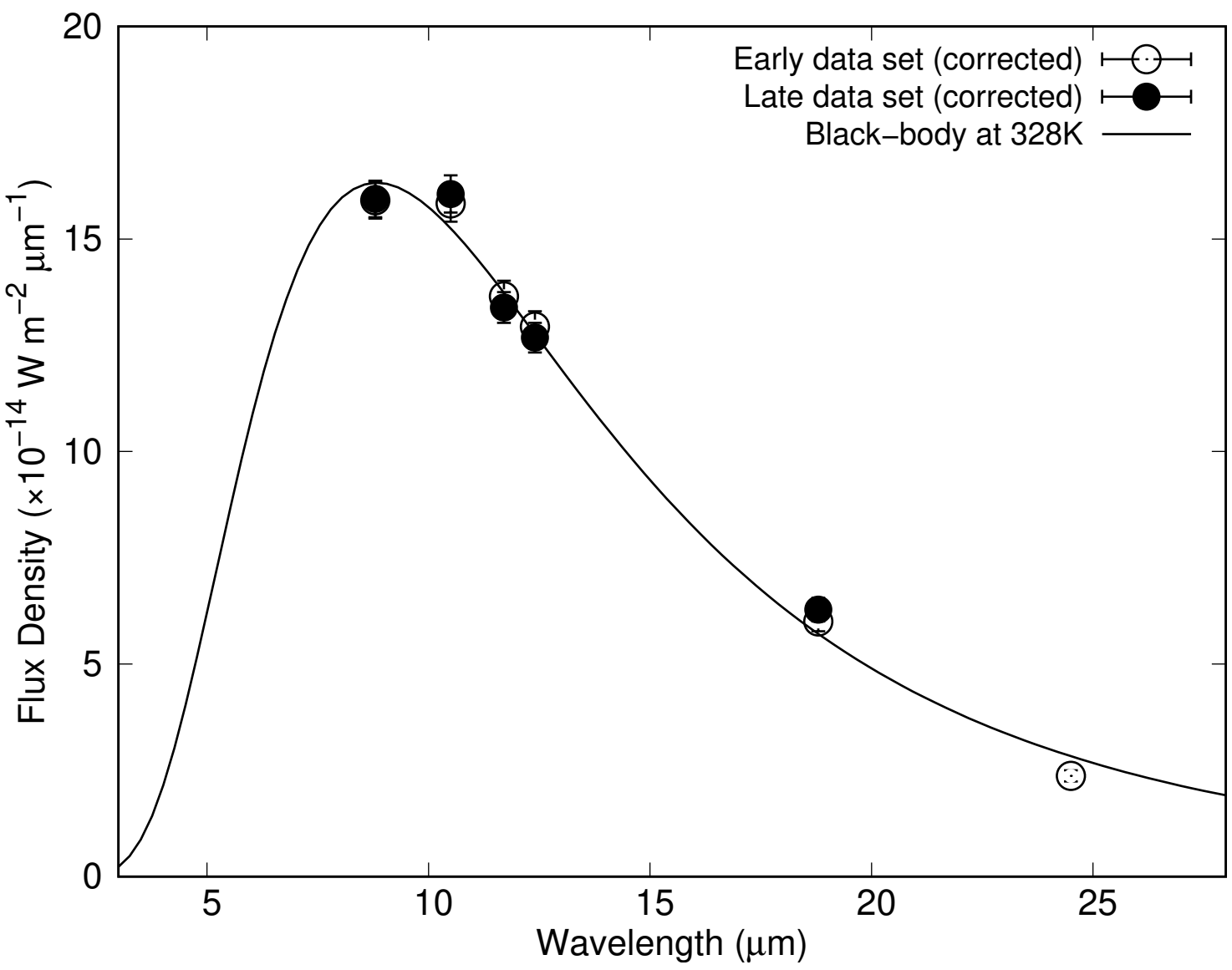

Figure 5: Spectral energy distribution of 2P/Encke based on the photometric data, corrected to the epoch of the *N*-band spectroscopic observation (UT 2003 November 12.29). The solid line is the best-fit blackbody spectrum (at 328 K) for our observations.

## 3.2 Spectroscopic data

Figure 6 shows the low-resolution spectrum of 2P/Encke in *N*-band obtained by the COMICS observations. The spectrum was flux-calibrated and corrected for the slit-loss, based on the photometric data at 8.8 and 12.4 $\mu$m (corrected to the epoch of the spectroscopic observation, see Figure 5). The observed spectrum displays no significant features, with the exception of a tentative absorption feature at ~9.5 μm. Although the 10-$\mu$m emission feature caused by cometary silicate grains in coma is usually seen in *N*-band spectra of comets (e.g., Y. Shinnaka et al. 2018), no clear feature is found in our spectrum. As discussed in the previous subsection based on the spatial profiles of 2P/Encke's images, it is likely that the spectral signals (collected within the rectangular aperture; $0.''33 \times 0.''83$) are dominated by the thermal radiation of the cometary nucleus.

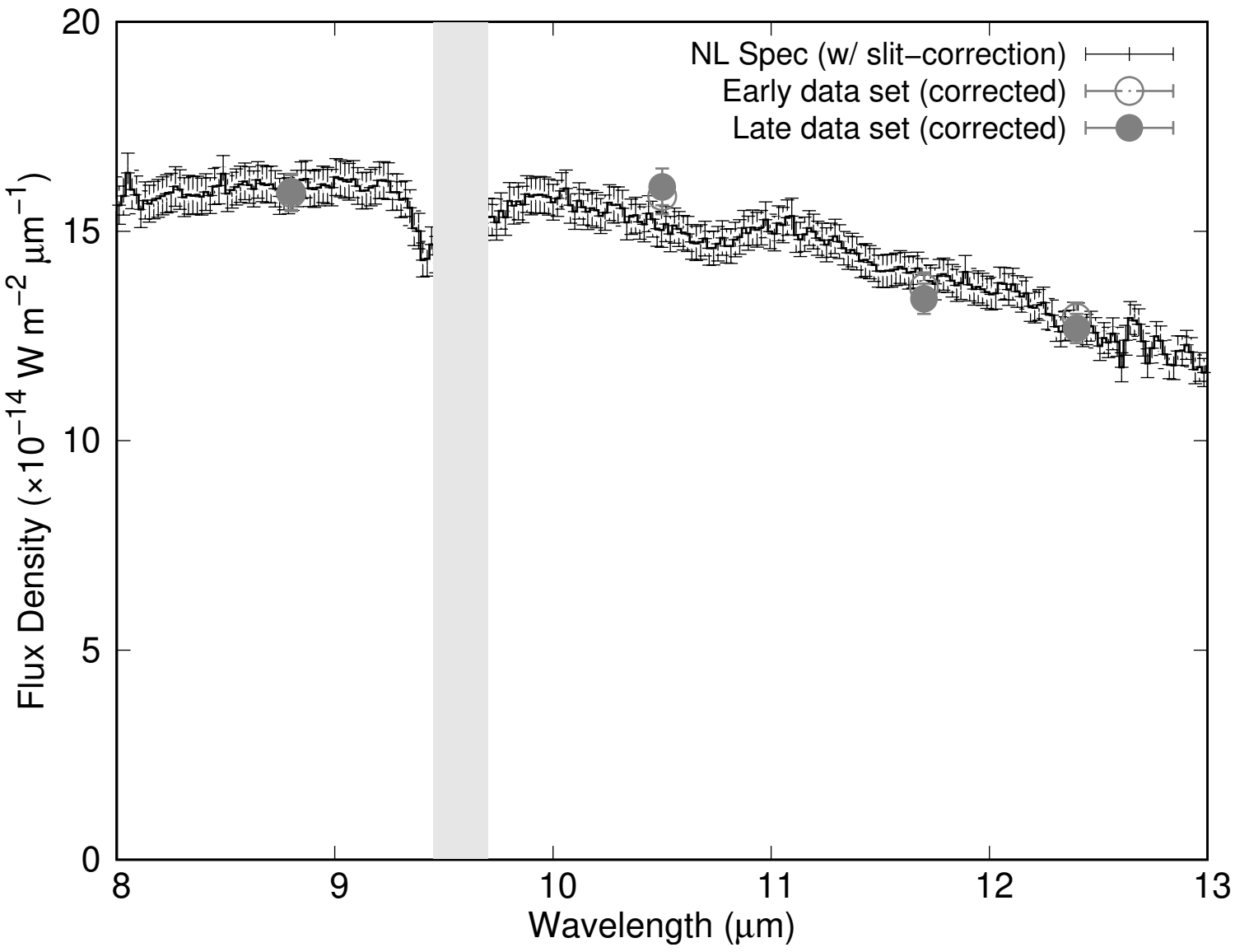

Figure 6: N-band low resolution spectrum of 2P/Encke on UT 2003 November 12.29, together with the photometric data shown in Figure 5. The grey-hatched area is the wavelength region influenced by the telluric ozone absorption.

4. Discussion and Conclusions

We estimated the semi-axial ratio of 2P/Encke as $a/b = 2.26^{+0.33}_{-0.21}$ by assuming the aspect angle $\alpha = 71.7°$ (K. Jockers et al. 2011 pole) as described in Section 3. These values are similar to the semi-axial ratio $a/b = 2.04$ derived by M. J. S. Belton et al. (2005) and consistent with the lower limit $a/b \geq 1.44 \pm 0.06$ obtained by S. C. Lowry and P. R. Weissman (2007). Our estimate is somewhat smaller than the semi-axial ratio $a/b \geq 2.6$ reported by Y. R. Fernandez et al. (2000) from mid-infrared thermal modeling, although their value is model-dependent. It is also consistent with the radar-based constraint by J. K. Harmon & M. C. Nolan (2005), who derived a long-axis dimension of ~9 km for an 11.1-h rotation period (corresponding to $a/b \approx 2.0 - 2.6$ depending on the assumed effective radius).

Regarding the surface temperature of the nucleus of 2P/Encke, the best-fit blackbody temperature is estimated as 328 ± 7 K for the photometric data on UT 2003 November 12. The surface temperature of a cometary nucleus is empirically estimated by $T_{\rm s} = 339.95\, r_{\rm h}{}^{-0.438}$ [K] for $r_{\rm h} = 0.8 \sim 3.4$ au; $r_{\rm h}$ is the heliocentric distance of comet (M. D. Suttle et al. 2020). Our result for 2P/Encke ($r_{\rm h} = 1.09$ au) is consistent with the empirical estimate for a comet at $r_{\rm h} = 1.09$ au, 327 K. However, the wavelength for the radiation peak is nearly at the shorter edge of our observations (Figure 5) and we need photometric data points at shorter wavelengths to estimate the temperature of nucleus surface more accurately.

We also refer to the flux densities at 4.49 and 5.73 $\mu$m on 2003 November 11 at 17:33 UT by the IRAC onboard the Spitzer Space Telescope (M. S. Kelley et al. 2006). We have converted their results (observed from the space telescope) to the values from the Earth with the $\Delta^{-2}$ law; $\Delta$ is the distance from 2P/Encke to the observer ($\Delta = 0.232$ au from the Spitzer Space Telescope). Their imaging observations at different wavelengths were performed simultaneously. Although they also reported the flux density at 3.55 $\mu$m for 2P/Encke, we did not use this result because it may have been contaminated by the reflected sunlight as illustrated by Figure 7 in M. S. Kelley et al. (2006). The photometric results at 4.49 and 5.73 $\mu$m are also corrected for the rotational phase based on the same manner as described in the previous section. The aspect angle from the Spitzer Space Telescope was $\alpha = 73.8°$, slightly different from that for the COMICS observations. We adopt our semi-axial ratio ($a/b = 2.26$), but we also consider the different semi-axial ratios, 2.04–2.6 (Y. R. Fernandez et al. 2000; M. J. S. Belton et al. 2005), to estimate the uncertainty of the rotational phase correction for the IRAC observations. Figure 7 shows the spectral energy distribution of 2P/Encke, including the photometric data obtained by the IRAC onboard the Spitzer Space Telescope. The best-fit blackbody temperature is 337±8 K, slightly higher than the value estimated from the COMICS observations only (but consistent with each other). The photometric data points by the IRAC observations seem to be systematically larger, probably because of the uncertainties in the rotational phase correction. Such discrepancy might be caused by the difference in the viewing geometry (i.e., line-of-sight vector) between the Spitzer Space Telescope and the Subaru Telescope, which were not strictly identical even with comparable aspect angles at the observations. Although we assumed a prolate, isothermal surface to estimate thermal radiation from the cometary nucleus, non-uniform temperature distribution on the cometary surface (e.g., the temperature gradient between day- and night-sides) might be important to convert the IRAC results to the observational conditions at the COMICS observations. Furthermore, we assumed a simple rotation (with a single spin axis) of a prolate body. An asymmetric ellipsoidal body and an excited rotational mode might also be important for 2P/Encke (M.J.S. Belton et al. 2005).

Finally, we discuss the spectroscopic features of the low-resolution N-band spectrum. T. Ootsubo et al. (2021) reported the absorption-like features in the *N*-band emissivity spectrum of the nucleus of comet 460P/PanSTARRS. We also divided the observed *N*-band spectrum of 2P/Encke by the blackbody spectrum at 328 K to compare with the emissivity spectrum of 460P/PanSTARRS. Figure 8 shows the relative emissivity spectra of these comets. Although they are not similar to each other, the absorption-like feature at ~9.5 $\mu$m (T. Ootsubo et al. 2021) is recognized in both spectra. The depth of this feature in 2P/Encke might be slightly influenced by the telluric ozone. We still do not know what the carrier of this feature is. We also do not know the variation of thermal properties of the materials covering cometary nuclei. More spectra of the cometary nucleus surfaces in the mid-infrared wavelength region are needed to identify the carriers of these features. The data will then be examined for the presence of various minerals, particularly hydrous silicates and organics (T. Ootsubo et al. 2021). For further discussion, we need simultaneous observations at different wavelengths or spectroscopy over a wider wavelength range than our observations. Future spectroscopic observations of 2P/Encke in mid-infrared by JWST/MIRI (G. S. Wright et al. 2023) will enable us to measure the surface temperature of the nucleus of this comet more precisely.

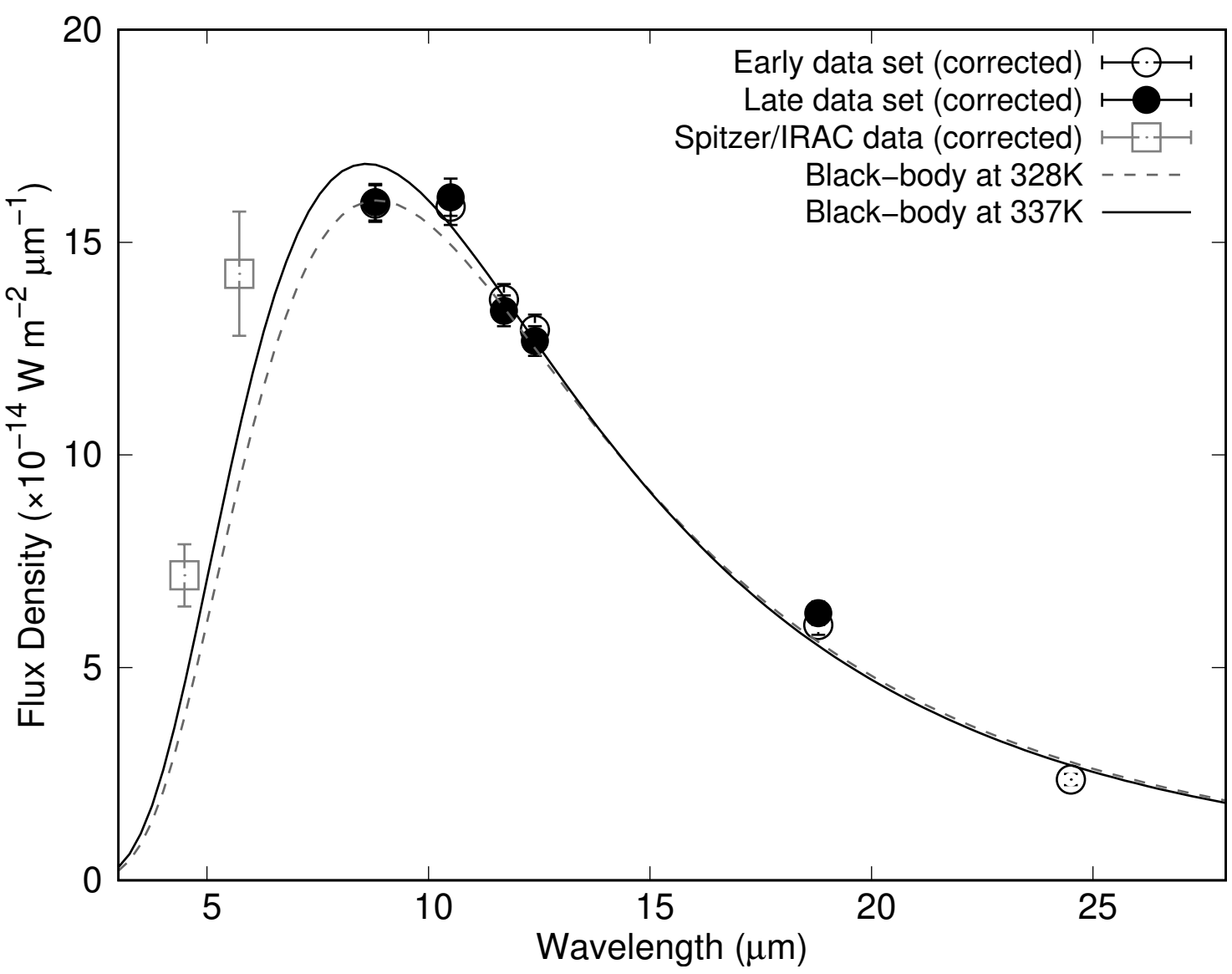

Figure 7: Spectral energy distribution of 2P/Encke based on the photometric data of both the COMICS observations and the IRAC observations, corrected to the epoch of the *N*-band spectroscopic observation by COMICS (UT 2003 November 12.29). The solid line is the best-fit blackbody spectrum (at 336 K) for all observations. The dashed line is the best-fit blackbody spectrum (at 328 K) for the COMICS observations only.

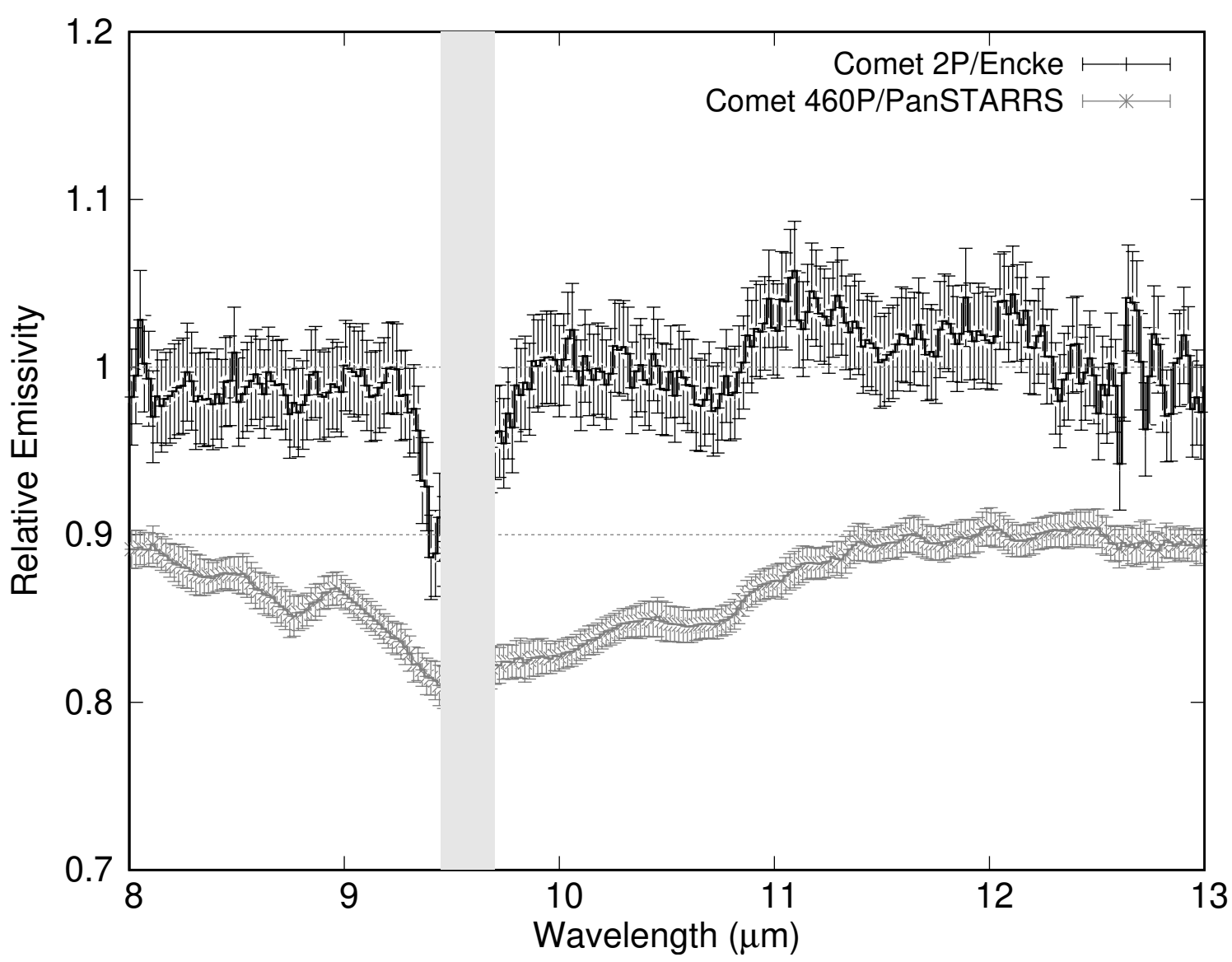

Figure 8: Relative emissivity spectrum of 2P/Encke's nucleus obtained by the COMICS observations, in comparison with that of 460P/PanSTARRS (T. Ootsubo et al. 2021). The spectrum of 460P/PanSTARRS (in grey) is shifted by 0.1 for the readability. The grey hatched area is the wavelength region influenced by the telluric ozone absorption.

As a summary, we conclude that

- we observed comet 2P/Encke in the mid-infrared wavelength region were performed on UT 2003 November 12.3, with COMICS on the Subaru telescope. The comet was at 1.09 au from the Sun and 0.27 au from the observer,
- the photometric observations were carried out with narrow-band filters centered at 8.8, 10.5, 11.7, 12.4, 18.8, and 24.5 $\mu$m. The low-resolution spectroscopic observations were performed in *N*-band (8–13 $\mu$m, R~250),
- the observed comet signals were dominated by the thermal radiation from 2P/Encke's nucleus, thanks to high-spatial resolution of COMICS,
- we estimated the semi-axial ratio of the nucleus as $a/b = 2.26^{+0.33}_{-0.21}$ based on the prolate ellipsoid model, and also estimated the surface temperature as 328 ± 7 K from the photometric data,
- the observed low-resolution *N*-band spectrum of 2P/Encke's nucleus showed an overall smooth continuum with a negative feature at ~9.5 $\mu$m.

Future spectroscopic observations of 2P/Encke in mid-infrared by JWST/MIRI, for wider wavelength range, are strongly encouraged because they will enable us to measure the surface temperature and the thermal emissivity spectrum (for the surface materials) of 2P/Encke's nucleus more precisely.

---

Appendix A

The aspect angle $\alpha$, defined as the angle between the nucleus spin axis and the comet-to-Earth direction (J. K. Harmon & M. C. Nolan 2005), was computed from the geocentric position of the comet and published pole directions. The spin-axis direction is a fixed direction on the celestial sphere that does not depend on the observer's location, while the geocentric coordinates of the comet represent the Earth-to-comet direction. Since the comet-to-Earth direction is the antipodal direction, the aspect angle is obtained as $\alpha = 180° - \theta$, where $\theta$ is the angular separation between the pole direction $(\mathrm{RA_p}, \mathrm{Dec_p})$ and the geocentric position of the comet $(\mathrm{RA_c}, \mathrm{Dec_c})$.

To compute $\theta$, we convert both directions to Cartesian unit vectors:

$$\boldsymbol{p} = (\cos \mathrm{Dec_p} \cos \mathrm{RA_p}, \ \cos \mathrm{Dec_p} \sin \mathrm{RA_p}, \ \sin \mathrm{Dec_p}),$$

$$\boldsymbol{c} = (\cos \mathrm{Dec_c} \cos \mathrm{RA_c}, \ \cos \mathrm{Dec_c} \sin \mathrm{RA_c}, \ \sin \mathrm{Dec_c}),$$

where RA and Dec are in radians. The angular separation is then given by

$$\cos\theta = \boldsymbol{p} \cdot \boldsymbol{c} = \cos \mathrm{Dec_p} \cos \mathrm{Dec_c} \cos(\mathrm{RA_p} - \mathrm{RA_c}) + \sin \mathrm{Dec_p} \sin \mathrm{Dec_c},$$

and the aspect angle $\alpha$ is

$$\alpha = 180° - \theta = 180° - \arccos(\boldsymbol{p} \cdot \boldsymbol{c}).$$

Acknowledgement

The authors also wish to recognize and acknowledge the very significant cultural role and reverence that the summit of Maunakea has always had within the indigenous Hawaiian community. This research was financially supported by Japan Society for the Promotion Science (Grant No. JP15J10864, JP20K14541, JP21H04498, JP23K25930, and JP24H00271) and by Koyama Space Science Institute of Kyoto Sangyo University.

REFERENCES

M. J. S. Belton, N. H. Samarasinha, Y. R. Fernández, & K. J. Meech 2005, Icarus, 175, 181–193.

M. Cohen, R. G. Walker, B. Carter, et al., 1999, AJ, 117, 1864–1889.

Y. R. Fernández, C. M. Lisse, K. H. Ulrich, et al. 2000, Icarus, 147, 145–160.

Y. R. Fernández, S. C. Lowry, P. R. Weissman, B. E. A. Mueller, et al. 2005, Icarus, 175, 194–214.

M. C. Festou & O. Barale 2000, AJ, 119, 3119–3132.

J. K. Harmon & M. C. Nolan 2005, Icarus, 176, 175–183.

K. Jockers, S. Szutowicz, G. Villanueva, T. Bonev, & P. Hartoph 2011, Icarus, 215, 153–185.

H. Kataza, Y. Okamoto, S. Takubo, et al., 2000, in Proc. SPIE 4008, 1144–1152.

M. S. Kelley, C. E. Woodward, D. E. Harker, et al. 2006, ApJ, 651, 1256–1271.

M. M. Knight, R. Kokotanekova, & N. H. Samarasinha 2024, in COMETS III (K. J. Meech, et al. eds.), pp. 407–432. Univ. of Arizona, Tucson.

P. L. Lamy, I. Toth, Y. R. Fernández, H. A. Weaver 2004, in COMETS II (M. C. Festou, et al. eds.), pp. 223–264. Univ. of Arizona, Tucson.

H. F. Levison & L. Dones 2014, in Encyclopedia of the Solar System (T. Spohn, D. Breuer and T. V. Johnson, eds.), pp. 705–719. Elsevier Inc.

M. Lippi, G. L. Villanueva, M. J. Mumma, & S. Faggi 2021, AJ, 162, 74.

S. C. Lowry & P. R. Weissman 2007, Icarus, 188, 212–223.

C. G. Mason, R. D. Gehrz, T. J. Jones, et al. 2001, ApJ, 549, 635–646.

Y. K. Okamoto, H. Kataza, T. Yamashita, et al., 2003, in Proc. SPIE 4841, 169–180.

T. Ootsubo, H. Kawakita, & Y. Shinnaka 2021, Icarus, 363, id.114425.

T. Ootsubo, H. Kawakita, T. Takata, et al. 2026, Publications of the Astronomical Society of Japan, id.psag088.

Y. L. Radeva, M. J. Mumma, G. L. Villanueva, et al. 2013, Icarus, 223, 298–307.

Z. Sekanina 1988a, AJ, 95, 911.

Z. Sekanina 1988b, AJ, 96, 1455.

Z. Sekanina 1991, Journal of the Royal Astronomical Society of Canada, 85, 324–376.

Y. Shinnaka, T. Ootsubo, H. Kawakita, M. Yamaguchi, M. Honda, & J. Watanabe 2018, AJ, 156, id.242.

M. D. Suttle, L. Folco, M. J. Genge, & S. S. Russell 2020, Icarus 351, id.113956.

D. Tody 1986, in Proc. SPIE Instrumentation in Astronomy VI (D. L. Crawford, ed.), 627, 733.

D. Tody 1993, in Astronomical Data Analysis Software and Systems II, A.S.P. Conf. Ser., Vol. 52 (R. J. Hanisch, R.J.V. Brissenden, & J. Barnes, eds.), 173.

G. S. Wright, G. H. Rieke, A. Glasse, et al. 2023, PASP, 135, 1046, id.048003.